\documentclass[%
 reprint,
superscriptaddress,
 amsmath,amssymb,
 aps,
 pra,
]{revtex4-2}

\usepackage{amsmath}
\usepackage{graphicx}% Include figure files
\usepackage{dcolumn}% Align table columns on decimal point
\usepackage{bm}% bold math
\usepackage{xcolor}
\usepackage[version=4]{mhchem}
\usepackage{xr}
\usepackage{multirow}
\usepackage{array}
\usepackage[T1]{fontenc}
\usepackage[utf8]{inputenc}
\usepackage[title]{appendix}
\usepackage[normalem]{ulem}
\usepackage{soul}
\soulregister\cite7
\usepackage[colorlinks=true,citecolor=blue,linkcolor=red,linktocpage=true]{hyperref}

\newcommand{\bp}{\mathbf{p}}
\newcommand{\cW}{\mathcal{W}}

\newcolumntype{P}[1]{>{\centering\arraybackslash}p{#1}}

\definecolor{myorange}{RGB}{245,156,74}
\definecolor{mygreen}{RGB}{17,159,87}

\newcommand{\rev}[1]{\textcolor{blue}{#1}}

\begin{document}

\preprint{APS/123-QED}

\title{Topological edge currents promote exploratory search in microtubule dynamic instability}

\author{Chongbin Zheng}

\affiliation{
 Department of Physics and Astronomy, Rice University, Houston, Texas 77005, USA 
}
\affiliation{
 Center for Theoretical Biological Physics, Rice University, Houston, Texas 77005, USA
}

\author{Jaime Agudo-Canalejo}

\affiliation{
Department of Physics and Astronomy, University College London, London WC1E 6BT, United Kingdom 
}

\author{Jonathon Howard}

\affiliation{
Department of Molecular Biophysics and Biochemistry, Yale University, New Haven, Connecticut 06511, USA
}

\author{Evelyn Tang}
\thanks{Email: e.tang@rice.edu}

\affiliation{
 Department of Physics and Astronomy, Rice University, Houston, Texas 77005, USA
}
\affiliation{
 Center for Theoretical Biological Physics, Rice University, Houston, Texas 77005, USA
}

\begin{abstract}
Microtubules capture chromosomes during mitosis by stochastically switching between growth and shrinkage at catastrophe events. They display strikingly rich biochemistry and dynamics, regulated by a stabilizing cap with distinct conformational states. Microtubule lengths at catastrophe are observed to follow a peaked distribution, while their growth ``stutters'' briefly before catastrophe. Such complexity makes it hard to capture all these observations without a large number of tunable parameters. Here, we introduce a topological model of the microtubule cap that reproduces the features above through dynamical edge states, that provides a minimal description with just two free parameters. Our approach further provides an analytical description of catastrophes and allows the same features to persist over a wide range of tubulin concentration, consistent with experimental observations.

\end{abstract}

\maketitle

\section{Introduction}
Exploratory processes are common in biology, such as in the various processes of target search \cite{mattila2008filopodia,andrew2007chemotaxis,gerhart2007theory,nalbant2018exploratory,sims2008scaling}. One notable example is in the process of mitosis, when chromosomes are divided into two exact copies for each daughter cell \cite{alberts2022molecular}. This process relies on microtubule filaments that search for and capture chromosomes \cite{kirschner1986beyond,heald2015thirty}, and subsequently pull the two copies to opposite ends of the dividing cell \cite{mcintosh2016mitosis}. Curiously, microtubules grow and shrink repeatedly during target search, often growing to a very different length each time before shrinking at so-called catastrophe events \cite{gudimchuk2021regulation}. These dynamics of the microtubules were first discovered by Mitchison and Kirschner in the 1980s, and dubbed dynamic instability \cite{mitchison1984dynamic}.

\begin{figure*}[ht!]
	\centering
	\includegraphics[width=17cm, height=14.7cm]{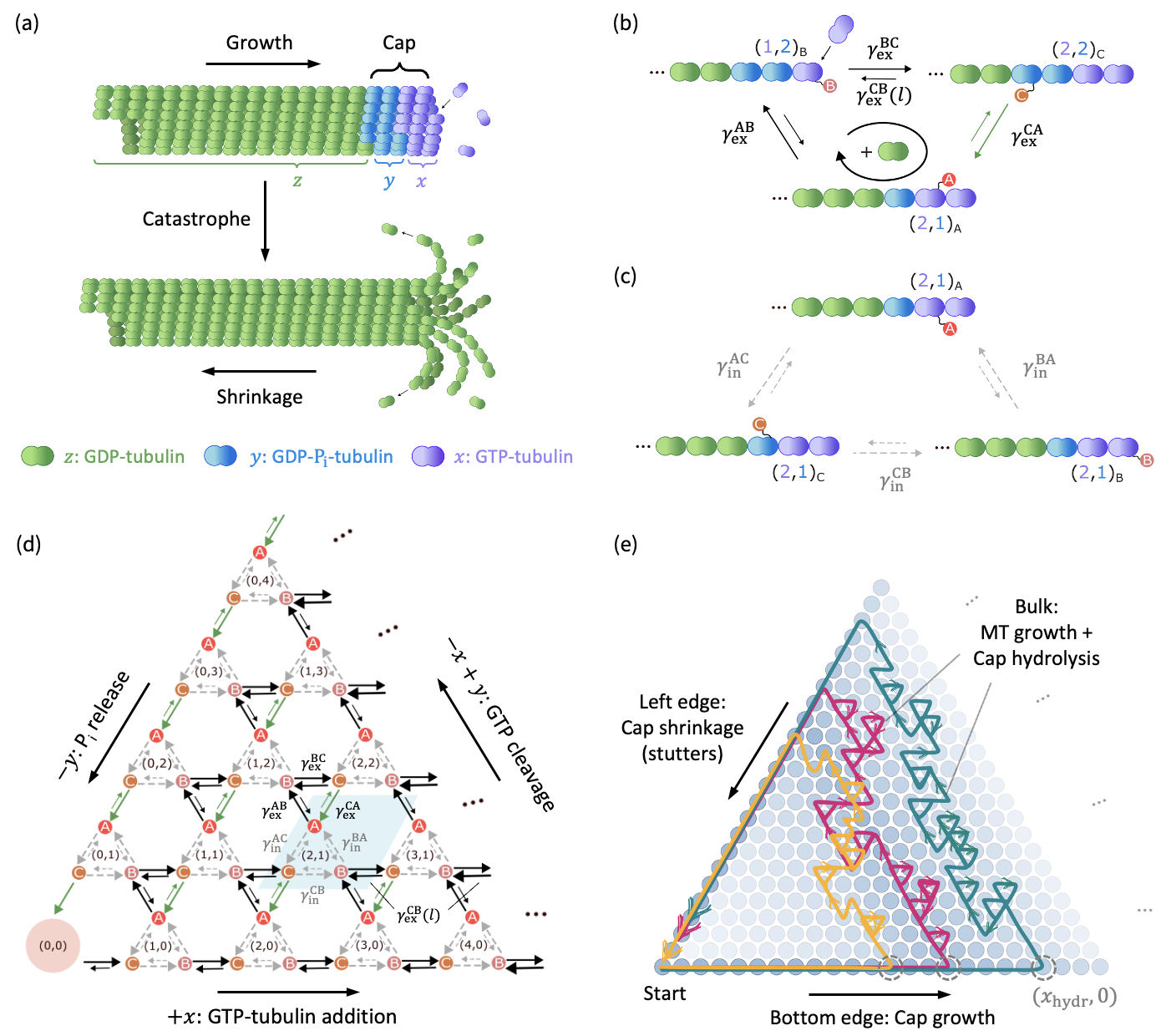}
	\caption{Topological model for microtubule dynamics. (a) Microtubules grow in the presence of a stabilizing cap, made from GTP-tubulin (purple) and GDP-P$_\text{i}$-tubulin dimers (blue). Cap loss leads to catastrophe events, followed by rapid microtubule shrinkage. (b) The cap is modified by cyclic external transitions (solid arrows): GTP-tubulin addition (top), GTP cleavage (left), and P$_\text{i}$ release (right). Brackets $(x,y)$ record the number of GTP-tubulin and GDP-P$_\text{i}$-tubulin dimers, which change with transitions shown by black arrows. Green arrows denote reactions that change the number of GDP-tubulin (recorded separately). After one clockwise cycle, the cap remains unchanged but the microtubule grows by one GDP-tubulin dimer. (c) For each $(x,y)$, there are three internal conformational states (A, B, C) that transition through dashed arrows; each internal state primes the cap for a different external reaction. (d) Repeating the reaction cycles along the $x$ and $y$ axes forms a Kagome lattice; the repeated motif is highlighted in blue. The GTP-tubulin dissociation rate $\gamma_\text{ex}^\text{CB}$ can increase with cap length (horizontal axis). (e) Schematic of three stochastic trajectories on a background where darker circles represent sites that are visited more frequently. Catastrophes proceed via cap growth along the bottom edge, followed by two-step hydrolysis through the bulk and left edge. Hydrolysis starts at the encircled points denoted by $(x_\text{hydr},0)$.}
    \label{fig_model}
\end{figure*} 

Microtubules display strikingly rich structure and biochemistry, where a stabilizing cap with distinct conformational and nucleotide states \cite{brouhard2018microtubule} regulates dynamic instability \cite{gudimchuk2020mechanisms}. Dynamic instability shows complex dynamics, where microtubule lengths at catastrophe follow a peaked distribution \cite{odde1995kinetics,gardner2011depolymerizing} and undergo transient ``stutters'' (paused growth) right before catastrophes \cite{maurer2014eb1,mahserejian2022quantification} -- over a large range of tubulin concentrations. Yet, this richness and complexity is difficult to model theoretically and analytically. Existing models capture at most one of the above observations and often rely on many tunable parameters. For instance, simple models that describe catastrophe as a single-step process cannot reproduce a peaked distribution \cite{verde1992control,dogterom1993physical}. Meanwhile, more complex models capture the cap structure in detail \cite{coombes2013evolving,zakharov2015molecular,alexandrova2022theory} and some features such as the peaked distribution \cite{bowne2013microtubule,coombes2013evolving,li2014theoretical,zakharov2015molecular,alexandrova2022theory}, stutters \cite{kim2019long}, or microtubule growth and shrinkage rates \cite{zakharov2015molecular,kim2019long,alexandrova2022theory}. However, no existing model simultaneously shows the key conditions for dynamic instability, the peaked catastrophe distribution and stutters, let alone a minimal description that offers mechanistic insight and analytical results with few fitting parameters.

A novel approach to modeling dynamical biochemical processes that has recently emerged is topology. Topological states produce dimensional reduction such as global cycles \cite{tang2021topology,zheng2024topological} or multistable fixed points \cite{nelson2025topology} within large  networks of biochemical reactions. Such topological models were first developed in quantum systems \cite{moore2010birth,chiu2016classification,stormer1999fractional,tang2012interacting} and later extended to classical systems like mechanical lattices \cite{kane2014topological,susstrunk2015observation}, photonics \cite{xiao2014surface,pocock2018topological,benalcazar2017quantized,weimann2017topologically}, and electrical circuits \cite{imhof2018topolectrical,hofmann2019chiral}. In the context of biological systems, topological models have been used to describe the circadian rhythm \cite{zheng2024topological}, sensory adaptation during chemotaxis \cite{murugan2017topologically,dasbiswas2018topological}, and gene transcription networks \cite{nelson2025topology}. Crucially, the dynamics localize to the edge of a system in a topological phase in a way that is robust to defects and disorder \cite{agudo2025topological}. This framework has also been applied to microtubules, to capture the large range of catastrophe lengths \cite{tang2021topology} or emergent vibrational or electronic modes that destabilize filament growth \cite{prodan2009topological,aslam2019experimentally,subramanyan2023microtubules}. However, these models only account for very specific features and have not been closely compared to experimental data. For example, microtubules in Ref. \cite{tang2021topology} shrink from the minus end and mostly at the same catastrophe length, contrary to experimental observations \cite{mitchison1984dynamic,walker1988dynamic,gardner2011depolymerizing}.

In this work, we propose a model of the microtubule cap that quantitatively reproduces the peaked catastrophe length distribution \cite{gardner2011rapid} and generates the observed stutters \cite{maurer2014eb1,mahserejian2022quantification}. In our model, the onset of catastrophes and stutters arises from a topological mechanism, which provides a new paradigm for understanding microtubule dynamics. In particular, the cap dynamics are described by edge currents that follow lower-dimensional trajectories in the topological phase. This dimensional reduction allows these key features to be captured with only two free parameters, and further provides an analytical description of when catastrophes occur. In addition, the topological dynamics are robust to changes in reaction rates, enabling our model to reproduce the same features across a wide range of tubulin concentration.

\section{Topological model for microtubule cap} \label{sec_model}
\subsection{Model setup}
We introduce a stochastic model of the microtubule cap, which plays a central role in regulating microtubule dynamics. Microtubules are cylindrical polymer structures assembled from multiple protofilaments of tubulin dimers \cite{desai1997microtubule}, as illustrated in Fig. \ref{fig_model}(a). The cap is located at the microtubule plus end \cite{mitchison1984dynamic,gardner2011depolymerizing}, shown in purple and blue on the top right of the figure. It enables microtubule growth and stabilizes the microtubule under tubulin addition \cite{hyman1992role} while its removal immediately triggers catastrophes, i.e., rapid shrinkage from the same end \cite{walker1989asymmetric,bowne2013microtubule} (bottom of Fig. \ref{fig_model}(a)). Based on experimental evidence for a structurally distinct intermediate state in the cap \cite{maurer2014eb1,zhang2015mechanistic,manka2018role}, we model the cap with two components: GTP-tubulin dimers (purple) are followed by GDP-P$_\text{i}$-tubulin dimers (blue), a hydrolysis intermediate from chemical cleavage of phosphate (P$_\text{i}$) \cite{melki1990direct,melki1996continuous}. As hydrolysis proceeds, GTP-tubulin is eventually converted into unstable GDP-tubulin (green), which makes up the rest of the filament. 

Following previous theoretical work \cite{flyvbjerg1994stochastic,margolin2006analysis,padinhateeri2012random}, we use a simplified description where the microtubule is treated as a single-protofilament polymer. The numbers of GTP-tubulin, GDP-P$_\text{i}$-tubulin, and GDP-tubulin dimers in the protofilament are denoted by $x,y,z$ respectively. We assume that catastrophes occur when the cap is entirely lost at $x=y=0$, based on experimental evidence that a single layer of GTP-tubulin is sufficient to stabilize microtubules \cite{drechsel1994minimum,caplow1996evidence}. Upon catastrophe, the microtubule depolymerizes completely, and its total length $x+y+z$ shrinks to zero before growth resumes.

During microtubule growth, the number of dimers in the cap, denoted by $(x,y)$, changes through biochemical reactions illustrated in Fig. \ref{fig_model}(b). GTP-tubulin dimers add to or dissociate from the protofilament at the plus end (horizontal arrows that change $x$). Once added, GTP-tubulin undergoes hydrolysis in two steps: GTP cleavage converts GTP-tubulin to GDP-P$_\text{i}$-tubulin (bottom to top-left arrow that change $x$ to $y$), while P$_\text{i}$ release converts GDP-P$_\text{i}$-tubulin to GDP-tubulin (top-right to bottom arrow that change $y$ to $z$). One clockwise cycle of these transitions leave the cap state $(x,y)$ unchanged, while the microtubule grows by one GDP-tubulin dimer (tracked by $z$). In Fig. \ref{fig_model}(b), we label P$_\text{i}$ release and its reverse reaction by green arrows as they modify the GDP-tubulin dimers (in green) outside the cap. As we treat the microtubule as a single protofilament, reactions in Fig. \ref{fig_model}(b) take coarse-grained rates averaged over all protofilaments.

Each reaction above can be primed, e.g., by conformational changes or elastic stress that propagate across the whole cap through allostery \cite{brouhard2018microtubule}. Such long-range interactions can arise from conformational mismatch between tubulin dimers with different curvatures and nucleotide states, and are increasingly recognized as essential for understanding catastrophes \cite{kim2019long,igaev2020microtubule}. For example, a conformational expansion at the interface between GDP-P$_\text{i}$-tubulin and GDP-tubulin could prime the microtubule for P$_\text{i}$ release \cite{estevez2020structural}. Separately, Refs. \cite{prodan2009topological,aslam2019experimentally,subramanyan2023microtubules} demonstrated topological phonon modes at the microtubule tip, which could disrupt GTP-tubulin addition and prime GTP cleavage.

In our model, we propose three distinct ``internal'' states for the cap, each priming one of the ``external'' transitions in Fig. \ref{fig_model}(b) (GTP-tubulin association, GTP cleavage, P$_\text{i}$ release). This is supported by structural evidence of three distinct tubulin conformational states \cite{brouhard2018microtubule}. We represent these states with the internal coordinate s (subscript A, B, or C), so that the full molecular state space of the cap is defined with $(x,y)_\text{s}$. For a topological model, three internal states are also the minimum to capture the non-equilibrium cycles of microtubule growth and catastrophe, which is driven thermodynamically by GTP hydrolysis \cite{desai1997microtubule}. In Sec. \ref{sec_1D}, we consider a variant of our model with two internal states, which fails to reproduce experimental data. While more internal states could plausibly represent additional conformational intermediates, we consider a minimal model with three states, which is sufficient to capture existing experiments.

Besides external transitions that change the number of dimers $x$ and $y$, we also consider internal transitions that only change the conformational state s while keeping $x$ and $y$ fixed (Fig. \ref{fig_model}(c)). For example, the conformational expansion in Ref. \cite{estevez2020structural} converts state A (primed for GTP cleavage) to state C (primed for P$_\text{i}$ release). We denote external and internal transition rates by $\gamma_\text{ex}^{ij}$ and $\gamma_\text{in}^{ij}$, respectively, where $i$ and $j$ denote the initial and final conformational state of the transition with $i\neq j$.

Microtubules are strikingly out-of-equilibrium, where free energy released from GTP hydrolysis biases certain biochemical reactions (e.g. GTP cleavage or tubulin conformational changes) over their reverse transitions \cite{desai1997microtubule,brouhard2018microtubule}. Accordingly, we assign larger transition rates to these driven processes, as indicated by larger forward arrows in Fig. \ref{fig_model}(b) and \ref{fig_model}(c). We assume that these reactions happen at specific places in the cap, i.e., GTP cleavage only occurs at the interface between GTP-tubulin and GDP-P$_\text{i}$-tubulin, and P$_\text{i}$ release only occurs at the interface between GDP-P$_\text{i}$-tubulin and GDP-tubulin. This is consistent with experimental findings of increased hydrolysis activity at such interfaces \cite{igaev2020microtubule}, and similar to the ``vectorial'' hydrolysis mechanism in previous models \cite{hill1984introductory,hinow2009continuous,ranjith2009nonequilibrium}.

The transitions in Fig. \ref{fig_model}(b) and Fig. \ref{fig_model}(c) form the basic motifs in our model. Repeating these motifs over possible cap states $(x,y)_\text{s}$ forms a Kagome lattice, which has the connectivity shown in Fig. \ref{fig_model}(d) (repeated motif highlighted in blue). A Kagome lattice state space has also been used in a previous model \cite{tang2021topology}, but it did not describe the microtubule cap or contained reactions realistic to microtubules. In contrast, our model maps specific biophysical reactions in the microtubule cap onto this lattice. In particular, the edges of the lattice correspond to cap states with the minimum or maximum number of GTP-tubulin or GDP-P$_\text{i}$-tubulin dimers. For example, on the bottom edge the cap contains only GTP-tubulin and no GDP-P$_\text{i}$-tubulin, while on the left edge the cap contains only GDP-P$_\text{i}$-tubulin where all GTP have been cleaved. 

The connectivity of the lattice reflects our assumption that each cap conformation only allows a specific set of reactions to occur, which leads to transitions between neighboring states rather than all-to-all connectivity. Note that this state space only describes the cap dynamics. We assume that catastrophes occur when the cap is entirely lost, which happens when the cap returns to state $(0,0)$ after traversing the state space. Cap loss is assumed to be irreversible, as indicated by a single arrow from $(0,1)_\text{C}$ to $(0,0)$ in Fig. \ref{fig_model}(d).

\subsection{Model dynamics}
Our model supports two distinct dynamical regimes, determined by the competition between external and internal transitions from the same state $(x,y)_\text{s}$. As shown in Ref. \cite{tang2021topology}, when external transitions such as tubulin addition or hydrolysis dominate ($\gamma_\text{ex}^{ij}>\gamma_\text{in}^{ik}$), the state space supports persistent edge currents: once the system encounters the edge it will likely continue along the edge, as can be seen by inspection. The emergence of such edge currents depends on the global pattern of transition rates in state space, which can be captured by a nontrivial topological invariant (also see Appendix \ref{appendix_invariant}) \cite{tang2021topology,agudo2025topological}. We therefore refer to this regime as the topological phase. Conversely, when internal transitions are faster ($\gamma_\text{ex}^{ij}\lesssim\gamma_\text{in}^{ik}$), the system is in a trivial phase and undergoes random growth and hydrolysis throughout state space via diffusive motion, rather than directed motion along the edges.

Since microtubule growth and hydrolysis ($\gamma_\text{ex}^{ij}$) are highly non-equilibrium processes \cite{desai1997microtubule,brouhard2018microtubule} compared to internal conformational changes ($\gamma_\text{in}^{ik}$), we expect $\gamma_\text{ex}^{ij}>\gamma_\text{in}^{ik}$ and focus on the topological phase. We further allow the GTP-tubulin dissociation rate $\gamma_\text{ex}^{\text{CB}}$ to increase as the cap length $l=x+y$ grows, e.g., $\gamma_\text{ex}^{\text{CB}}\propto l^n$ for $n>0$. This dependence is motivated by observations of a tubulin conformational gradient \cite{roostalu2020speed} and increasing curvature \cite{mcintosh2018microtubules} towards the end of the cap, where conformational mismatch between neighboring dimers weaken tubulin interactions and lead to faster dissociation for longer caps. The weakened interactions may also come from a more ``tapered'' cap structure, where terminal tubulin dimers have fewer lateral neighbors to interact with \cite{gardner2011rapid,coombes2013evolving}. Meanwhile, accumulation of structural defects (e.g., missing dimers or changing protofilament number) can further destabilize longer caps \cite{schaedel2019lattice,bowne2013microtubule}. In our single-protofilament description, these mechanisms are captured phenomenologically by the rate $\gamma_\text{ex}^{\text{CB}}(l)$, which is coarse-grained over all protofilaments to account for the overall effects of structural changes in the cap. Such structural changes can be resolved in more detail in a full multi-protofilament description, which we outline in Sec. \ref{sec_discussion}. In our model, this length dependence allows for a peaked catastrophe length distribution consistent with \cite{gardner2011depolymerizing}, as we will show in Sec. \ref{sec_validation}.

The increasing dissociation rate makes GTP-tubulin addition increasingly unlikely, introducing a ``soft'' boundary as opposed to a ``hard'' boundary with a limited state space \cite{tang2021topology}. This soft boundary separates the topological and trivial phases. To the right of the soft boundary at large cap lengths, the system is in a trivial phase without edge states. To the left, the system is in a topological phase, where edge currents describe cap growth along the bottom edge and cap shrinkage along the left edge (Fig. \ref{fig_model}(e). Moreover, microtubule growth is paused along the left edge, as no GTP-tubulin addition occurs there. These pauses match experimental observations of brief stutters before catastrophes \cite{maurer2014eb1,mahserejian2022quantification}, which are captured naturally by left edge currents in our model. 

In between the trivial and the topological phase, the soft boundary permits a mixture of edge and diffusive dynamics, where stochastic trajectories go through directed diffusion from the bottom to the left edge. This corresponds to cap hydrolysis from GTP-tubulin to GDP-P$_\text{i}$-tubulin and growth of the microtubule length $x+y+z$. We illustrate three representative trajectories in Fig. \ref{fig_model}(e) with different colors, where darker circles in the background represent states that are visited more frequently. These dynamics qualitatively capture the different phases of growth, stutter, and shrinkage in dynamic instability. In particular, catastrophes involve a sequence of edge and bulk dynamics in the 2D state space: the cap grows along the bottom edge, followed by a two-step hydrolysis process in the bulk and on the left edge. To further reproduce experimental results quantitatively, we match the transition rates more closely to experimental data below.

\begin{figure}[ht!]
	\centering
	\includegraphics[width=9cm, height=13cm]{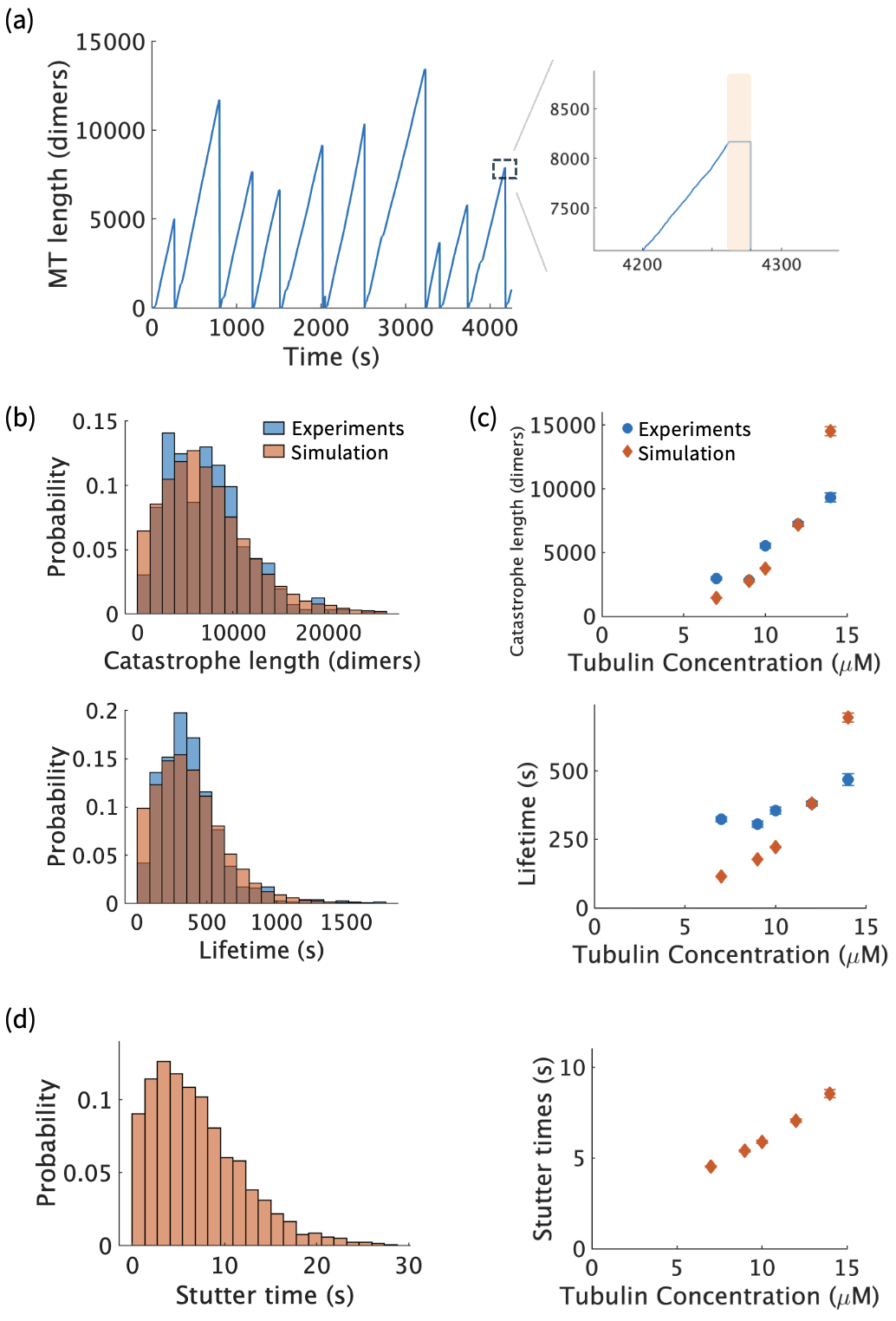}
	\caption{The topological model reproduces key features of catastrophe observed in experiments. (a) Microtubule length as a function of time, from model simulations. Catastrophes occur over a large range of length scales. Right: microtubule growth stutters briefly before catastrophe (yellow shaded region). (b) Distribution of catastrophe length and microtubule lifetime, fitted to experimental data from Ref \cite{gardner2011depolymerizing} at $12\mu$M tubulin. Our model generates peaked distributions for both quantities. (c) Average catastrophe length and lifetime for different tubulin concentrations. Consistent with experiments from \cite{gardner2011depolymerizing}, both quantities increase with concentration. Error bars represent one standard error. (d) Left: Distribution of stutter times from model simulations, which also shows a peak. Right: Average stutter time increases with tubulin concentration.}
	\label{fig_results}
\end{figure}

\section{Validation with experimental data} \label{sec_validation}
To compare our model behavior with experimental data, we perform stochastic simulations using the Gillespie algorithm \cite{gillespie1977exact} (details in Appendix \ref{appendix_simulations}). Our simulations focus on the regime $\gamma_\text{ex}^{ij}>\gamma_\text{in}^{ik}$, which allows for the edge dynamics shown in Fig. \ref{fig_model}(e). A representative time trace of microtubule length generated by our simulations is shown in Fig. \ref{fig_results}(a), where microtubules grow at a nearly constant speed consistent with measurements \cite{walker1988dynamic,gardner2011depolymerizing}. The growth pauses briefly before each catastrophe, as highlighted on the right of Fig. \ref{fig_results}(a), capturing the experimentally observed stutters \cite{maurer2014eb1,mahserejian2022quantification}. In addition, the catastrophe lengths span over one order of magnitude, reproducing the broad range of length scales observed \cite{walker1988dynamic,fygenson1994phase,desai1997microtubule}.

To quantitatively connect our model to experiments, we define the transition rates using biophysical parameters from the literature. In principle, each transition rate in Fig. \ref{fig_model}(d) can be different, yielding as many as 12 parameters: a global scaling factor that sets the overall timescale, and 11 ratios between transition rates. We determine the global scaling by tuning a single base rate $\gamma_\text{ex}^{\text{BC}}$, while other rates scale proportionally based on the ratios. Three out of the 11 ratios can be constrained by previous experimental measurements. Where no experimental data are available, we simplify the model by taking certain ratios equal to the constrained ratios, as explained below. Such simplifications reduce the number of free parameters to two, which we fit to observed catastrophe length distributions \cite{gardner2011depolymerizing}. 

\begin{table*}[ht!]
    \renewcommand{\arraystretch}{1.3}
    \centering
    \begin{tabular}{P{2.3cm}|P{7cm}|P{2.2cm}|P{2cm}}
    \hline
    Transition rate & Description & Ratio to $\gamma_\text{ex}^\text{BC}$ & Value (s$^{-1}$) \\ \hline
    $\gamma_\text{ex}^\text{BC}$ & GTP-tubulin association rate at $12\mu$M tubulin & 1 & 37.7 \\ 
    $\gamma_\text{ex}^\text{AB}$ & GTP cleavage rate & $s_\text{st}$ & 245 \\ 
    $\gamma_\text{ex}^\text{CA}$ & P$_\text{i}$ release rate & $s_\text{st}$ & 245 \\ 
    $\gamma_\text{in}^\text{BA}$ & & $1/r$ & 0.31 \\ 
    $\gamma_\text{in}^\text{AC}$ & & $s_\text{st}/r$ & 2.04 \\ 
    $\gamma_\text{in}^\text{CB}$ & & $s_\text{st}/r_\text{P}$ & 2.45 \\ 
    $\gamma_\text{ex}^\text{CB}(l=1)$ & GTP-tubulin dissociation rate at $l=1$ & $K_\text{d}$ & 4.08 \\ 
    $\gamma_\text{ex}^\text{BA}$ & GTP synthesis rate & $s_\text{st}e^{-\mu}$ & 4.48 \\
    $\gamma_\text{ex}^\text{AC}$ & P$_\text{i}$ binding rate & $s_\text{st}e^{-\mu}$ & 4.48 \\
    $\gamma_\text{in}^\text{AB}$ & & $e^{-\mu}/r$ & 0.0057 \\
    $\gamma_\text{in}^\text{CA}$ & & $s_\text{st}e^{-\mu}$/r & 0.037 \\
    $\gamma_\text{in}^\text{BC}$ & & $s_\text{st}e^{-\mu}/r_\text{P}$ & 0.045 \\
    \hline
    \end{tabular}
    \caption{Transition rates at $12\mu$M tubulin and cap length $l=1$. The ratios $K_\text{d}=0.11$ and $s_\text{st}=9.5$ are constrained by experiments on microtubule growth \cite{wieczorek2015microtubule} and stutters \cite{maurer2014eb1,duellberg2016size,strothman2019microtubule}. The free parameters $r=120$ and $r_\text{P}=100$ are obtained from fitting the catastrophe length distribution in \cite{gardner2011depolymerizing}. Reverse transition rates (except GTP-tubulin dissociation) are related to their corresponding forward rates by the local detailed balance condition $\gamma_\text{ex}^{ji}/\gamma_\text{ex}^{ij}=\gamma_\text{in}^{ji}/\gamma_\text{in}^{ij}=e^{-\mu}$, with $\mu=4$ (see main text).}
    \label{table_parameters}
\end{table*}

We begin by defining the GTP-tubulin dissociation rate through the ratio $K_\text{d}=\gamma_\text{ex}^{\text{CB}}(l=1)/\gamma_\text{ex}^{\text{BC}}$, which characterizes GTP-tubulin binding affinity. Experimental measurements report that the ratio between the dissociation and apparent association rate constants is 1.3 $\mu$M \cite{wieczorek2015microtubule}. This quantity corresponds to $K_\text{d}[\text{Tub}]$ in our model, where [Tub] denotes the free tubulin concentration. We evaluate $K_\text{d}$ at a fixed concentration $[\text{Tub}]=12$ $\mu$M, the same as the catastrophe length data considered later \cite{gardner2011depolymerizing}. This yields $K_\text{d} = 1.3/12\approx0.11$. Single-molecule experiments have reported a smaller $K_\text{d}$ \cite{mickolajczyk2019direct}, but the qualitative behavior of our model do not change under specific choices of $K_\text{d}$ within the experimentally reported range. We further need to specify how the dissociation rate depends on cap length. For now, we assume a simple linear relationship $\gamma_\text{ex}^{\text{CB}}\propto l$, and discuss more general length dependencies in Sec. \ref{sec_analytical}.

Next, we define the GTP cleavage and P$_\text{i}$ release rates by $s_\text{st} = \gamma_\text{ex}^{\text{CA}}/\gamma_\text{ex}^{\text{BC}} = \gamma_\text{ex}^{\text{AB}}/\gamma_\text{ex}^{\text{BC}}$. In the absence of direct experimental measurements, we assume that the two rates are equal. The ratio $s_\text{st}$ scales the rate $\gamma_\text{ex}^{\text{CA}}$ and also $\gamma_\text{in}^{\text{AC}}$ defined below, which together control the duration of stutters. We tune $s_\text{st}$ to match experimental measurements of average stutter times at $\sim 7$ seconds \cite{maurer2014eb1,duellberg2016size,strothman2019microtubule}. Since $s_\text{st}$ only defines relative rates and not absolute stutter times, we tune it together with the global scaling factor (discussed later) that sets the absolute timescale, yielding $s_\text{st}=9.5$.

To capture the out-of-equilibrium nature of the transitions, we define the slower reverse rates $\gamma_\text{ex}^{ji}$ and $\gamma_\text{in}^{ji}$ by the local detailed balance condition $e^\mu=\gamma_\text{ex}^{ij}/\gamma_\text{ex}^{ji}=\gamma_\text{in}^{ij}/\gamma_\text{in}^{ji}$, where $\mu$ (in units of $k_\text{B} T$) measures the energy input that drives the system out of equilibrium \cite{hill1989}. Thus, $\mu$ sets the ratio for all pairs of external and internal transitions except GTP-tubulin association and dissociation, whose ratio is instead given by $K_\text{d}[\text{Tub}]$ discussed previously. For microtubules, the non-equilibrium driving comes from free energy released from GTP hydrolysis (GTP cleavage and P$_\text{i}$ release). Studies have found that the energy is not released at once during these two steps, but also stored in the microtubule lattice via elastic strain and drives tubulin conformational changes \cite{brouhard2018microtubule,kim2019long,stewman2020dynamic}. Since GTP hydrolysis releases $\sim20$ $k_\text{B}T$ of free energy per molecule \cite{brouhard2018microtubule,milo2015cell}, we set $\mu=20/5=4$, assuming that the non-equilibrium driving is equally distributed across the five reaction pairs (GTP cleavage, P$_\text{i}$ release, and the three internal conformational changes).

To finally obtain the length distributions, we specify the topology of the model by two ratios: $r=\gamma_\text{ex}^{\text{AB}}/\gamma_\text{in}^{\text{AC}}=\gamma_\text{ex}^{\text{BC}}/\gamma_\text{in}^{\text{BA}}$ (set equal for simplicity) and $r_\text{P}=\gamma_\text{ex}^{\text{CA}}/\gamma_\text{in}^{\text{CB}}$. These ratios measure the competition between forward external and internal transitions. As we have not found experimental data that constrains these ratios, we take them as free parameters to fit the experimentally observed catastrophe length distribution \cite{gardner2011depolymerizing}.

When fitting these parameters, we focus on the regime where external transitions are faster than internal transitions ($r>1$ and $r_\text{P}> 1$), which produces the edge dynamics in Fig. \ref{fig_model}(e). In this regime, $r$ and $r_\text{P}$ control the length of edge currents and the tendency to move towards the left edge during directed diffusion. To determine the best fit, we sweep through $(r,r_\text{P})$ and simulate 20000 catastrophe events for each parameter combination. The resulting catastrophe length distribution is compared to experimental data at 12 $\mu$M tubulin \cite{gardner2011depolymerizing} using the two-sample Kolmogorov-Smirnov (KS) statistic. The statistic is minimized at $r=120$ and $r_\text{P}=100$ (KS statistic = $0.04$, p-value = 0.21), with a 95\% bootstrap confidence interval of $r\in[95,130]$ and $r/r_\text{P}\in[1.14,1.36]$ (see Appendix \ref{appendix_simulations}). The low KS statistic and high p-value indicate no statistically significant difference between the distributions from simulation and experiments. The simulation results for these parameters are shown on the upper panel of Fig. \ref{fig_results}(b), which closely matches the peaked distribution observed \cite{gardner2011depolymerizing}. The quality of the fit does not depend sensitively on our specific choices of the constrained ratios: randomly perturbing these ratios and refitting $(r,r_\text{P})$ consistently yields p-values above 0.05 (see Appendix \ref{appendix_simulations}).

Furthermore, we set the overall timescale of the model by scaling the base rate $\gamma_\text{ex}^{\text{BC}}$ to match the average microtubule lifetime from experimental data \cite{gardner2011depolymerizing}. This yields $\gamma_\text{ex}^{\text{BC}}=37.7$ s$^{-1}$, consistent with single-molecule measurements of the GTP-tubulin association rate at $44.2\pm20.8$ s$^{-1}$ \cite{mickolajczyk2019direct}. Together with the 11 ratios from constraints and parameter fitting, this fully determines all transition rates in our model, which we summarize in Table \ref{table_parameters}. The resulting lifetime distribution is shown in the bottom panel of Fig. \ref{fig_results}(b), which is also peaked. The distribution qualitatively matches experimental results, although the quantitative agreement is weaker compared to the catastrophe length distribution (KS statistic = 0.06, p-value = 0.01).

We also compare our model behavior with experimental data at different tubulin concentrations [Tub]. To model changing concentrations, we set the GTP-tubulin association rate $\gamma_\text{ex}^\text{BC}$ proportional to [Tub] \cite{walker1988dynamic}. Interestingly, the GTP-tubulin dissociation rate was also found to increase linearly with [Tub] \cite{gardner2011rapid}. At higher [Tub], Ref. \cite{gardner2011rapid} showed that faster growth produces a more tapered tip structure, where terminal tubulin dimers on each protofilament have fewer lateral neighbors and dissociate more readily \cite{mickolajczyk2019direct}. In our single-protofilament description, we capture this changing structure by setting the coarse-grained dissociation rate as $\gamma_\text{ex}^\text{CB}(l=1,[\text{Tub}])=\gamma_\text{ex}^\text{CB,0}+\beta^\text{CB}[\text{Tub}]$. The constants $\gamma_\text{ex}^\text{CB,0}$ and $\beta^\text{CB}$ are fixed by two conditions. First, $\gamma_\text{ex}^\text{CB}$ at $l=1$ matches Table \ref{table_parameters} at $12\mu$M. Second, this dissociation rate equals the association rate $\gamma_\text{ex}^\text{BC}$ at $1\mu$M, the experimentally measured critical concentration where the two rates balance \cite{gardner2011rapid,wieczorek2015microtubule}.

Our model captures dynamic instability across a wide range of tubulin concentrations [Tub]. Past experiments reported dynamic instability for concentrations from 5 to 30 $\mu$M \cite{walker1988dynamic,fygenson1994phase,odde1995kinetics,gardner2011depolymerizing}. Across this broad range of concentrations, our model consistently reproduces dynamic instability in contrast to previous single-protofilament models \cite{margolin2006analysis,ranjith2009nonequilibrium,padinhateeri2012random,bowne2013microtubule}, as the topological edge currents remain robust to changing reaction rates from different concentrations.

Further, our model qualitatively captures how catastrophe length and microtubule lifetime change with tubulin concentration. Both quantities show an increasing trend with [Tub] in Fig. \ref{fig_results}(c), consistent with experimental data \cite{gardner2011depolymerizing}. Across the full range of concentrations considered, the model consistently generates peaked catastrophe length distributions, in agreement with experimental data \cite{gardner2011depolymerizing}. Moreover, these results do not change qualitatively even when the dissociation rate $\gamma_\text{ex}^\text{CB}$ is taken to be independent of [Tub]. These \rev{findings} highlight the generality of our model, which reproduces key experimental features \textit{in vitro} across different tubulin concentrations. Interestingly, the free tubulin concentration is actively buffered \textit{in vivo} and does not change as freely \cite{gasic2019autoregulation}.

In addition to capturing key experimental observations, our model also generates testable predictions on the duration of stutters. In our simulations, we measure the stutter time as the duration the system spends on the left edge of the state space before catastrophe. The stutter time follows a peaked distribution similar to catastrophe length and microtubule lifetime (Fig. \ref{fig_results}(d), left). Meanwhile, its average is expected to increase with the tubulin concentration, as shown on the right panel of Fig. \ref{fig_results}(d). 

\section{Analytical condition for peaked distribution} \label{sec_analytical}
\begin{figure}[t!]
	\centering
	\includegraphics[width=9cm, height=9.5cm]{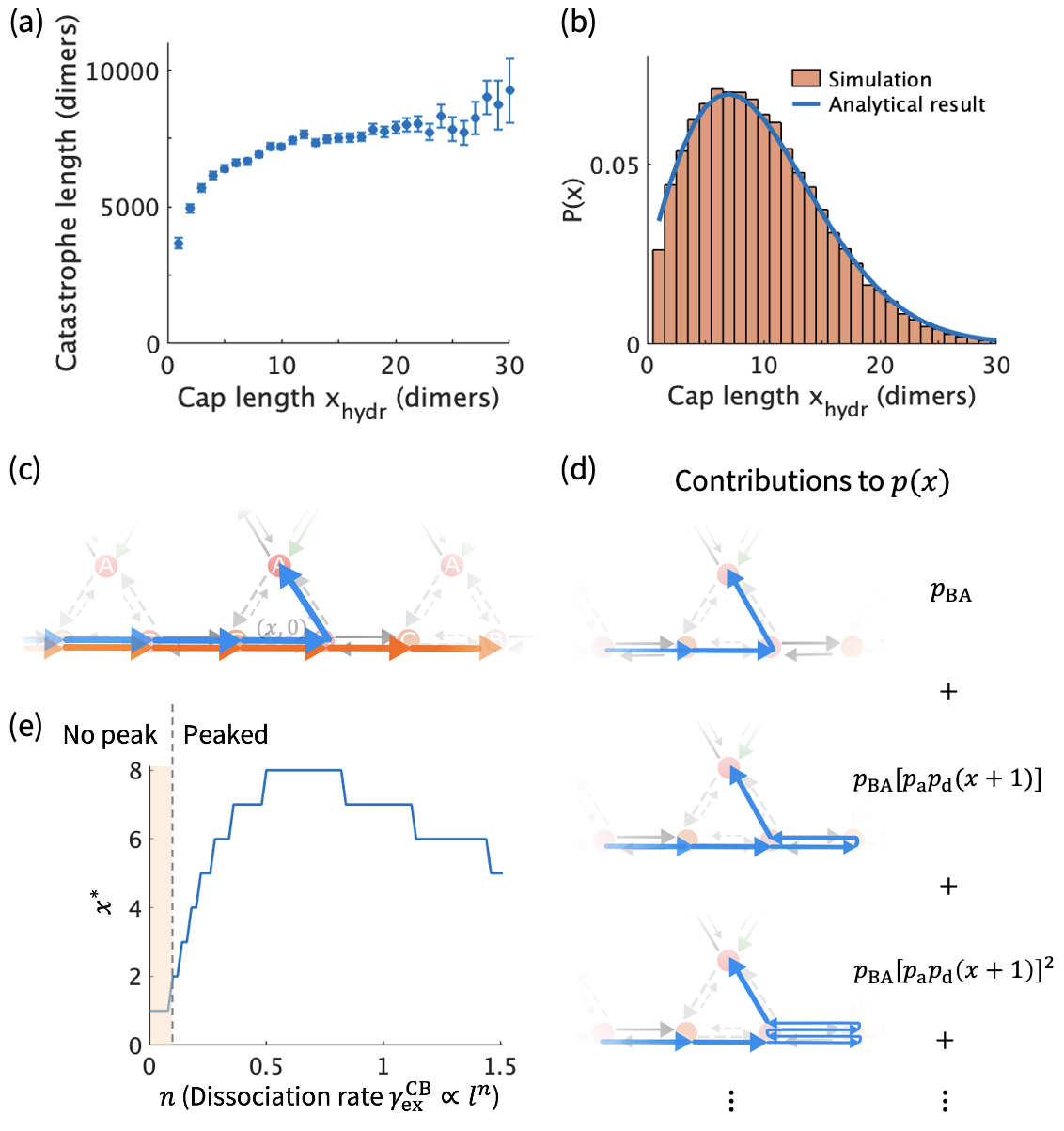}
	\caption{Bottom edge dynamics give an analytical condition for a peaked length distribution. (a) The cap length at the onset of hydrolysis, $x_\text{hydr}$, increases monotonically with catastrophe length. Error bars represent one standard error. (b) $P(x)$, the distribution for $x_\text{hydr}$, is peaked for our model parameters; here $r=120, r_\text{P}=100$. An analytical calculation (blue curve) agrees with simulation results. (c) On the bottom edge, a stochastic trajectory can either enter the bulk at $x$ (blue) or continue along the edge past $x$ (orange). (d) The bulk entry probability $p(x)$ takes contributions from a family of trajectories, which undergo different rounds of association/dissociation. The corresponding probabilities are shown on the right of each trajectory. (e) Our model provides an analytical condition for a peaked catastrophe length distribution, which arises when $P(x)$ has a maximum at $x^*>1$. This occurs when the dissociation rate $\gamma_\text{ex}^\text{CB}\propto l^n$ has exponent $n\geq 0.1$. Note that for $n<0.1$ (shaded area), $P(x)$ does not have a peak ($x^*=1$).}
	\label{fig_analytical}
\end{figure}

To understand when we expect a peaked catastrophe length distribution, we derive an analytical condition for when a peak emerges. Our analysis rests on a crucial simplification, where topological edge currents reduce the stochastic dynamics in our 2D state space to 1D trajectories along state space boundaries. As a result of this dimensional reduction, the total length of a trajectory depends on its length along the bottom edge, denoted by $x_\text{hydr}$ in Fig. \ref{fig_model}(e). The former governs the catastrophe length, while the latter corresponds to the cap length at the onset of cap hydrolysis. In our model, catastrophe length increases monotonically with $x_\text{hydr}$ (Fig. \ref{fig_analytical}(a)), which is intuitive because a longer cap leads to a more stable microtubule that grows for longer before catastrophe. As a result, the catastrophe length distribution is peaked whenever the $x_\text{hydr}$ distribution, denoted by $P(x)$, is peaked (e.g., see Fig. \ref{fig_analytical}(b)). This observation allows us to focus on the simpler distribution $P(x)$, where a peak emerges when the derivative $P'(x)=0$ at some positive $x=x^*$ that satisfies this condition.

We calculate $P(x)$ by analyzing stochastic trajectories along the bottom edge (Fig. \ref{fig_analytical}(c)). At each $x$ coordinate, a trajectory either enters the bulk of the state space (blue) with probability $p(x)$, or continues along the edge (orange) with probability $1-p(x)$, as we neglect higher-order effects such as less likely trajectories. $P(x)$ is then the probability that a trajectory continues along the edge at all previous $x-1$ steps, and enters the bulk at $x$, i.e.,
\begin{equation} \label{eqn_Px}
    P(x) = \left\{\prod\limits_{x'=1}^{x-1}\left[1-p(x')\right]\right\}p(x).
\end{equation}
This expression shows that length dependence is required for a peaked $P(x)$ distribution. If $p(x)$ is constant, $P(x)$ follows a geometric distribution and decays exponentially.

To obtain an analytical expression for $P(x)$, we express $p(x)$ in terms of transition rates in our model. $p(x)$ is given by the probability of a family of trajectories that enters the bulk at $x$, which involve different rounds of association and dissociation reactions between $x$ and $x+1$, as illustrated in Fig. \ref{fig_analytical}(d). To obtain the probability of such trajectories, we first define $p_\text{BA}$, the probability for the $(x,0)_\text{B}\rightarrow(x,0)_\text{A}$ internal transition:
\begin{equation} \label{eqn_pBA}
    p_\text{BA} = \frac{\gamma_{\text{in}}^{\text{BA}}}{\gamma_{\text{in}}^{\text{BA}} + \gamma^{\text{BC}}_{\text{in}} + \gamma_{\text{ex}}^{\text{BC}}}.
\end{equation}
Next, we define $p_\text{a}$ and $p_\text{d}(x)$, the probabilities for GTP-tubulin association or dissociation respectively, at cap length $l=x$:
\begin{align}
    p_\text{a} &= \frac{\gamma_{\text{ex}}^{\text{BC}}}{\gamma_{\text{in}}^{\text{BA}} + \gamma^{\text{BC}}_{\text{in}} + \gamma_{\text{ex}}^{\text{BC}}}, \\
    p_\text{d}(x) &= \frac{\gamma_{\text{ex}}^{\text{CB}}(x)}{\gamma^{\text{CA}}_{\text{in}} + \gamma_{\text{in}}^{\text{CB}} + \gamma_{\text{ex}}^{\text{CB}}(x)}. \label{eqn_pa}
\end{align} 
Each round of association and dissociation contributes a factor of $p_\text{a}p_\text{d}(x+1)$ to the trajectory probability, as shown in Fig. \ref{fig_analytical}(d). We neglect higher-order trajectories with two or more steps forward in $x$ before returning, as they go through backward internal transitions $\gamma_\text{in}^\text{BC}$ and have negligible probabilities.

Then $p(x)$ is given by the sum
\begin{align} \label{eqn_px}
\begin{split}
    p(x) =& p_\text{BA} \sum\limits_{m=0}^\infty \left[p_\text{a} p_\text{d}(x+1)\right]^m \\
    =& \frac{p_\text{BA}}{1-p_\text{a} p_\text{d}(x+1)}.
\end{split}
\end{align}
Substituting Equations (\ref{eqn_pBA}-\ref{eqn_px}) into Equation (\ref{eqn_Px}), we get the analytical expression for $P(x)$, which agrees with simulation results as shown by the blue curve in Fig. \ref{fig_analytical}(b). 

From the expression of $P(x)$, we obtain the analytical condition for a peak in both $P(x)$ and the catastrophe length distribution. Note that $P(x)$ is only defined at discrete values $x_\text{hydr}=1,2,...$, which represents the cap length at the onset of hydrolysis ($x=0$ is excluded, consistent with typical length measurements from experiments \cite{odde1995kinetics,gardner2011depolymerizing}). Therefore, $P(x)$ is peaked only when its maximum $P(x^*)$ occurs at $x^*>1$. This condition imposes a minimal rate at which the dissociation rate has to increase with $l$. For example, for $\gamma_{\text{ex}}^{\text{CB}}\propto l^n$, the rate has to increase as $l^{0.1}$ or more quickly in order to have a peak, as shown in Fig. \ref{fig_analytical}(e). This threshold $n^*$ can be calculated analytically by noting where $P'(x^*)=0$ and $x^*>1$ (see Appendix \ref{appendix_peaked}). It also remains concentrated around $n^*\sim0.1$ under random perturbations in model parameters (Appendix \ref{appendix_simulations}).

To conclude, we have shown that a cap-length-dependent dissociation rate $\gamma_{\text{ex}}^{\text{CB}}(l)$ is essential for generating a peaked distribution in our model. $\gamma_{\text{ex}}^{\text{CB}}(l)$ needs to increase with cap length sufficiently fast, but the required dependence can be different from the linear form we currently assume.

\begin{figure}[t]
	\centering
	\includegraphics[width=8.8cm, height=9.1cm]{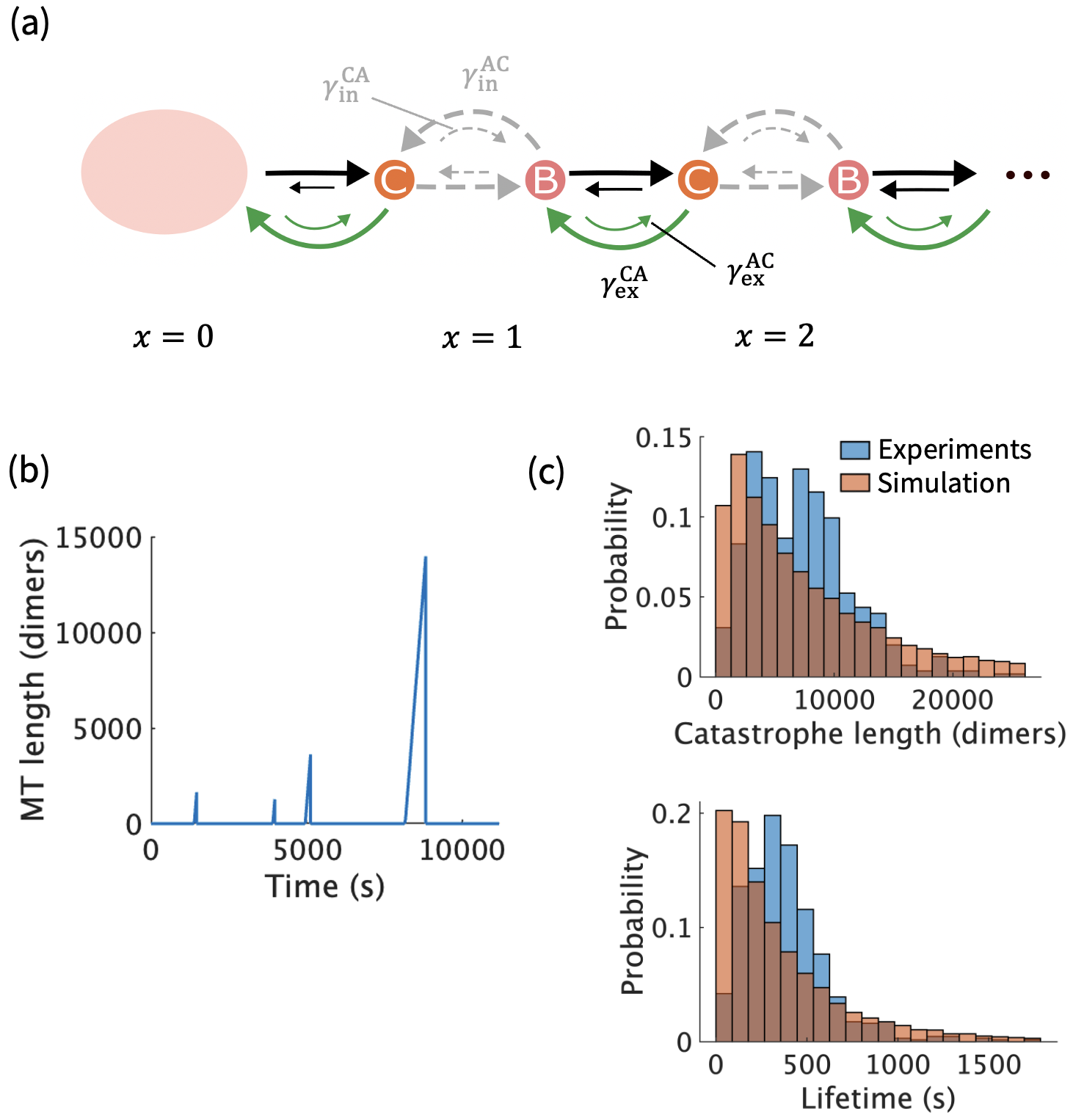}
	\caption{A reduced 1D model with a single-component cap fails to capture catastrophe behavior. (a) State space for the reduced 1D model, where GDP-P$_\text{i}$-tubulin is coarse-grained out. $x$ denotes the cap length. Curved arrows denote transitions that remain after coarse-graining the 2D model. (b) Microtubule length as a function of time, from the 1D model. Microtubules do not regrow immediately after catastrophe. (c) Distributions of catastrophe length and lifetime give poor agreement with the same experimental data compared to the 2D model.}
	\label{fig_1D}
\end{figure} 

\section{Single-component cap model fails to capture catastrophe behavior} \label{sec_1D}
Thus far, we have modeled the microtubule using a two-component cap. To investigate whether both degrees of freedom are necessary, we compare our model with a reduced single-component cap model. Following earlier work \cite{padinhateeri2012random,li2013theoretical,li2014theoretical}, we assume that GTP cleavage is fast relative to other timescales and coarse-grain out the reaction, treating GTP-tubulin and GDP-P$_\text{i}$-tubulin as the same species. This merges internal states A and B, reducing our molecular state space from a 2D Kagome lattice to a 1D chain with two internal states (Fig. \ref{fig_1D}(a)), where $x$ now denotes the cap length. The GTP-tubulin dissociation rate still depends on cap length as $\gamma_\text{ex}^\text{CB}\propto l$, as indicated by larger arrows to the right of Fig. \ref{fig_1D}(a).

The single-component cap model fails to reproduce the peaked length distribution and other key features of catastrophe. To compare with data, we refit the free parameters $r$, $r_\text{P}$ and the base rate $\gamma_\text{ex}^{\text{BC}}$ to the same catastrophe length and lifetime data from \cite{gardner2011depolymerizing}, keeping all other ratios fixed. Similar to Sec. \ref{sec_validation}, we minimize the Kolmogorov-Smirnov statistic, which yields $r=7.3\times10^4,r_\text{P}=6.0\times 10^4,\gamma_\text{ex}^{\text{BC}}=27.8$ s$^{-1}$ (KS statistic = 0.15). In this 1D model, microtubules remain short for extended periods after catastrophe (Fig. \ref{fig_1D}(b)), inconsistent with experimental observations \cite{walker1988dynamic,fygenson1994phase,gardner2011depolymerizing}. Moreover, growth transitions directly to shrinkage without a stutter phase. The resulting catastrophe length and lifetime distributions also show poor agreement with experimental data \cite{gardner2011depolymerizing}, as shown in Fig. \ref{fig_1D}(c). Such discrepancy with experiments shows that dynamics along the 1D edge is not sufficient to generate a peaked distribution, and points to the need for an extra degree of freedom. Our results are consistent with the failure of past single-protofilament models to capture the peaked catastrophe length distribution with only one cap degree of freedom \cite{flyvbjerg1994stochastic,flyvbjerg1996microtubule,margolin2006analysis}, unless additional constraints such as a multi-step catastrophe are imposed \cite{li2014theoretical,kok2025eb3}.

\section{Discussion} \label{sec_discussion}
In this study, we develop a topological model of the microtubule cap that captures three key features of dynamic instability: a dynamic cap structure \cite{brouhard2018microtubule,gudimchuk2021regulation}, a peaked catastrophe length distribution \cite{odde1995kinetics,gardner2011depolymerizing,alexandrova2022theory}, and transient stutters before catastrophe \cite{mahserejian2022quantification}. In contrast to more complex biochemical models \cite{coombes2013evolving,zakharov2015molecular,kim2019long,alexandrova2022theory}, we reproduce these experimental observations using a minimal description of cap dynamics with only two free parameters. Our work introduces a new paradigm for understanding dynamic instability in terms of topological edge currents: catastrophe lengths depend on the bottom edge currents, while stutters arise from left edge currents. The edge currents also result in dimensional reduction of the stochastic dynamics in the 2D state space, which enables an analytical description of catastrophes. The topological dynamics are robust to changes in system parameters, allowing dynamic instability and the peaked distribution to persist in a wide range of tubulin concentrations. 

Our model further identifies the conditions required to produce the peaked catastrophe length distribution. Analytical results show that a cap-length-dependent GTP-tubulin dissociation rate is needed, suggesting the role of increasing cap instability in regulating catastrophe dynamics. Moreover, we find that a two-component cap model, which incorporates two degrees of freedom via a two-step hydrolysis process, is needed to capture the experimental features, while a reduced single-component cap model with only one degree of freedom is insufficient. 

While our minimal single-protofilament model captures many key features of dynamic instability, it does not resolve the full structure of the microtubule cap. We have modeled the dissociation rate through a phenomenological cap-length dependence, rather than via underlying structural changes like lateral bonding and conformational mismatch between tubulin dimers \cite{manka2018role,kim2019long,mickolajczyk2019direct}. These effects can be captured explicitly in a multi-protofilament generalization, where the dissociation rate would depend on the number and nucleotide state of their neighbors, and potentially couple to more distant dimers due to long-range interactions \cite{kim2019long}. In the generalized model, each protofilament $i$ carries its own copy of the state space in Fig. \ref{fig_model}(d), labeled by $(x_i,y_i)_{\text{s}_i}$. We expect that topological edge currents would persist within the state space of individual protofilaments and remain important for microtubule dynamics, while how the edge dynamics interact with the inter-protofilament coupling remains an open question for future work.

The topological dynamics in our model yields several testable experimental predictions. First, we predict a delay of P$_\text{i}$ release following microtubule growth, where this delay lengthens for higher tubulin concentrations. In existing models \cite{melki1996continuous,coombes2013evolving,zakharov2015molecular,kim2019long,alexandrova2022theory}, such delays arise  from a finite rate of GTP cleavage preceding P$_\text{i}$ release, which does not depend on tubulin concentration. In contrast, P$_\text{i}$ release is delayed in our model because hydrolysis does not occur until trajectories leave the bottom edge. The delay corresponds to the time spent on this edge, which is expected to increase with tubulin concentration. Separately, we predict that the distribution of stutter times is peaked and the average stutter time increases with tubulin concentration [Tub] (see Fig. \ref{fig_results}(d)). This is because the stutter time in our model is determined by the cap length when trajectories reach the left edge, which follows a peaked distribution similar to $x_\text{hydr}$ and grows with [Tub]. 

By connecting to concrete biochemical reactions, we demonstrate how topology promotes microtubule exploration through protected edge currents in the molecular state space of the cap. More broadly, our work suggests a new mechanism that utilizes topological dynamics for exploratory behavior over a wide range of length scales. This topological mechanism can provide insight on other macromolecular exploratory structures \cite{kondev2025biological}, e.g., filopodia of neuronal  growth cones \cite{mattila2008filopodia,odde1995kinetics}, that similarly rely on cycles of growth and shrinkage for search and movement.

\section*{Acknowledgments}

We thank Dan Needleman, Jane Kondev and Rob Raphael for helpful comments. We gratefully acknowledge support from the NSF Center for Theoretical Biological Physics (PHY-2019745), the NSF CAREER Award (DMR-2238667), and the Chan-Zuckerburg Foundation. 

\section*{Author contributions}
All authors contributed to model development, analysis, and the writing of the manuscript.
\newline

\begin{appendices}
\section{Topological invariant} \label{appendix_invariant}
In this section, we introduce the topological invariant that characterizes the topology of our model, which is a non-Hermitian generalization of the invariant in Ref. \cite{ni2017topological}. First, we compute the invariant for an isotropic system with C$_3$-symmetric and constant transition rates throughout the lattice. We then show that our model with anisotropic rates lies in the same topological phase as the isotropic model when the rates are kept constant. Lastly, we reintroduce the cap-length-dependent GTP-tubulin dissociation rate $\gamma_\text{ex}^\text{CB}$ and discuss how it affects the topology of the model.

The dynamics of our model is described by the master equation
\begin{equation}
    \frac{\text{d}\bp}{\text{d}t}=\cW\bp,
\end{equation}
where $\bp(t)$ is the probability distribution over system states and $\cW$ is the transition-rate matrix. This linear equation is formally analogous to the Schrodinger equation $i\hbar\frac{\text{d}\Psi}{\text{d}t}=H\Psi$. Since the Hamiltonian $H$ only differs from $\cW$ by a constant factor $i\hbar$, they share the same eigenvectors. As we discuss below, the topological invariant we compute only depends on the eigenvectors, which is thus identical to that defined for the Hamiltonian in Ref. \cite{ni2017topological}. Our stochastic system also inherits the bulk-boundary correspondence in quantum systems as shown in \cite{tang2021topology}, so that a non-trivial invariant leads to the presence of a robust steady state localized on the edge of the state space.

Consider an isotropic system with the same state space as in Fig. \ref{fig_model}(d), but with C$_3$-symmetric reaction rates  $\gamma_\text{ex}^\text{BC}=\gamma_\text{ex}^\text{CA}=\gamma_\text{ex}^\text{AB}$ and $\gamma_\text{in}^\text{BA}=\gamma_\text{in}^\text{AC}=\gamma_\text{in}^\text{CB}$. We assume that their reverse rates $\gamma_\text{ex}^{ji}$ and $\gamma_\text{in}^{ji}$ are also isotropic, given by the local detailed balance condition $\gamma_\text{ex}^{ji}/\gamma_\text{ex}^{ij}=\gamma_\text{in}^{ji}/\gamma_\text{in}^{ij}=e^{-\mu}$. In periodic boundary conditions, the transition-rate matrix in momentum space is
\begin{widetext}
\begin{equation}
    \cW(\mathbf{k}) =
\begin{pmatrix}
    -(\gamma_\text{in}^\text{AB} + \gamma_\text{ex}^\text{AB}+\gamma_\text{in}^\text{AC} + \gamma_\text{ex}^\text{AC}) & \gamma_\text{in}^\text{BA} + \gamma_\text{ex}^\text{BA} e^{i (\mathbf{k}\cdot \mathbf{R_2})} & \gamma_\text{in}^\text{CA} + \gamma_\text{ex}^\text{CA} e^{-i (\mathbf{k}\cdot \mathbf{R_1})} \\
    \gamma_\text{in}^\text{AB} + \gamma_\text{ex}^\text{AB} e^{-i (\mathbf{k}\cdot \mathbf{R_2})} & -(\gamma_\text{in}^\text{BA} + \gamma_\text{ex}^\text{BA}+\gamma_\text{in}^\text{BC} + \gamma_\text{ex}^\text{BC}) & \gamma_\text{in}^\text{CB} + \gamma_\text{ex}^\text{CB} e^{-i k_x} \\
    \gamma_\text{in}^\text{AC} + \gamma_\text{ex}^\text{AC} e^{i (\mathbf{k}\cdot \mathbf{R_1})} & \gamma_\text{in}^\text{BC} + \gamma_\text{ex}^\text{BC} e^{i k_x} & -(\gamma_\text{in}^\text{CA} + \gamma_\text{ex}^\text{CA}+\gamma_\text{in}^\text{CB} + \gamma_\text{ex}^\text{CB})
\end{pmatrix},
\end{equation}
\end{widetext}
where $\mathbf{k}=(k_x,k_y)$, $\mathbf{R_1}=(\frac{1}{2},\frac{\sqrt{3}}{2})$, and $\mathbf{R_2}=(\frac{1}{2},-\frac{\sqrt{3}}{2})$. The topology of this system is set by the ratio of forward external and internal rates, which we define as 
\begin{equation}
    r_\text{iso} = \gamma_\text{ex}^\text{BC} / \gamma_\text{in}^\text{BA} = \gamma_\text{ex}^\text{CB} / \gamma_\text{in}^\text{AB}.
\end{equation}

To compute the invariant, we follow Ref. \cite{ni2017topological} and define the unitary transformation
\begin{equation} \label{eqn_W}
    U = \frac{1}{\sqrt{3}}
\begin{pmatrix}
    1 & 1 & 1 \\
    1 & e^{-\frac{2i\pi}{3}} & e^{\frac{2i\pi}{3}} \\
    1 & e^{\frac{2i\pi}{3}} & e^{-\frac{2i\pi}{3}}
\end{pmatrix}.
\end{equation}
which block-diagonalizes $\cW(\mathbf{k})$ at the high-symmetry K point, $\mathbf{K}=(\frac{4\pi}{3},0)$. We truncate the resulting matrix $\cW'=U\cW U^{-1}$ to two bands by only extracting the first and third row and column of $\cW'$, which corresponds to the pair of bands that are degenerate at K when $r_\text{iso}=1$. For the resulting $2\times2$ matrix $\cW_K(\mathbf{k})$, we calculate the Berry curvature 
\begin{align}
\begin{split}
    \Omega_i(\mathbf{k}) = i \bigl(\langle &\partial_{k_x} L_i(\mathbf{k})|\partial_{k_y} R_i(\mathbf{k})\rangle \\
    -& \langle \partial_{k_y}L_i(\mathbf{k})|\partial_{k_x} R_i(\mathbf{k})\rangle\bigr) ,
\end{split}
\end{align}
where $|L_i(\mathbf{k})\rangle$ and $|R_i(\mathbf{k})\rangle$ are left and right eigenvectors of the transition-rate matrix $\cW_K(\mathbf{k})$ for band $i$. Since $\cW(\mathbf{k})$ is non-Hermitian for $\mu\neq 0$, $|L_i(\mathbf{k})\rangle\neq|R_i(\mathbf{k})\rangle$ in general. Finally, integrating the Berry curvature over the Brillouin zone
\begin{equation}
    w_i = \int_\text{BZ} \Omega_i(\mathbf{k}) \text{d}^2 \mathbf{k},
\end{equation}
we obtain a generalized winding number for each band that characterize the topology of the system. We find numerically that the invariant is $(0,0)$ for $r_\text{iso} < 1$ and $(-1,+1)$ for $r_\text{iso} > 1$.

Starting from this isotropic reference point, we can construct a continuous path to the anisotropic rates of our model evaluated at fixed cap length $l=1$ (Table \ref{table_parameters}). Denote the transition matrix of the isotropic model by $\cW_\text{iso}(\mathbf{k})$ and that of our model by $\cW_\text{model}(\mathbf{k})$. For $\lambda\in[0,1]$, we then define the family of matrices
\begin{equation}
    \cW_\lambda(\mathbf{k}) = (1-\lambda)\cW_\text{iso}(\mathbf{k}) + \lambda\cW_\text{model}(\mathbf{k}),
\end{equation}
so that $\cW_0(\mathbf{k})=\cW_\text{iso}(\mathbf{k})$ and $\cW_1(\mathbf{k})=\cW_\text{model}(\mathbf{k})$. For every $\lambda$ in between, the linear combination $\cW_\lambda(\mathbf{k})$ is also a valid transition-rate matrix. The family of matrices $\cW_\lambda(\mathbf{k})$ thus defines a continuous path that connects the isotropic model to our model. We have verified numerically that the band gap of $\cW_\lambda(\mathbf{k})$ remains open for $\lambda\in[0,1]$. Since a topological transition requires the gap to close, this shows that our model (where $\gamma_\text{ex}^{ij}>\gamma_\text{in}^{mn}$ in general) lies in the same topological phase as the isotropic system with $r_\text{iso}>1$.

In our model, we also assume that the GTP-tubulin dissociation rate $\gamma_\text{ex}^\text{CB}$ increases with the cap length $l$. For small $l$, the system is in the same topological phase established above. However, at $l\sim 10$, $\gamma_\text{ex}^\text{CB}(l)$ grows to be comparable to its reverse rate $\gamma_\text{ex}^\text{BC}$. For stochastic systems, it has been shown that the absence of non-reciprocity generally leads to uniform steady states with trivial topology \cite{nelson2024nonreciprocity}. As a result, the edge dynamics are disrupted at large $l$ and the system enters a trivial phase.

\section{Stochastic simulations} 
\label{appendix_simulations}
To study microtubule behavior in our model, we run stochastic simulations using the Gillespie algorithm \cite{gillespie1977exact}. Throughout the simulations, we count the number of GDP-tubulin dimers $z$, which allows us to track the microtubule length $x+y+z$. Simulations start at $(0,0)$, where $z$ is initialized to 0. $z$ increases with each P$_\text{i}$ release reaction $\gamma_\text{ex}^\text{CA}$ and decreases with each reverse reaction $\gamma_\text{ex}^\text{AC}$. When $z=0$, we set $\gamma_\text{ex}^\text{AC}=0$ since there is no GDP to which P$_\text{i}$ can bind. A catastrophe event occurs when the system returns to $(0,0)$, at which point we immediately reset $z$ to 0. Note that there is no transition from $(0,0)$ to $(0,1)_\text{C}$ in Fig. \ref{fig_model}(d). We assume cap loss is irreversible and always leads to catastrophe.

To obtain the distributions in Fig. \ref{fig_results}(b), we generate catastrophe length distributions over a grid of $(r,r_\text{P})$ values in increments of 5. The best fit corresponds to the parameter combination that gives the minimum two-sample KS statistic between simulations and experimental data. In our model, catastrophe length is defined as the microtubule length $z$ when the system reaches $(0,0)$. For easier comparison, experimental length data in micrometers is converted to number of dimers, assuming that each dimer is 8 nanometers long \cite{desai1997microtubule}. Meanwhile, microtubule lifetime is defined as the time between consecutive catastrophe events. Before plotting, we remove catastrophe events with catastrophe lengths shorter than 416 dimers, the minimum length recorded in \cite{gardner2011depolymerizing}. This accounts for experimental difficulties of observing short catastrophe lengths. 

The same procedure is used to obtain the average values in Fig. \ref{fig_results}(c) and the distributions for the 1D model in Fig. \ref{fig_1D}(c). In Fig. \ref{fig_1D}(c), the peak in simulation results is an artifact of removing short catastrophe lengths. The raw simulation data for the 1D model does not have a peak, unlike the 2D model which is always peaked.

We further estimate confidence intervals for the fitting parameters $(r,r_\text{P})$ by bootstrapping. Specifically, we resample the experimental data with replacement to the same sample size, and minimize the KS statistic over the same grid of $(r,r_\text{P})$ values. The simulated catastrophe length distributions are held fixed for each combination of $(r,r_\text{P})$, while the resampling and the parameter optimization procedure are repeated for 2000 times. The 95\% confidence interval is then given by the 2.5th and 97.5th percentiles of the 2000 refitted parameter values. We find that $r$ and $r_\text{P}$ are strongly correlated across different trials (Pearson coefficient = 0.99), and thus report confidence intervals in $r$ and $r/r_\text{P}$ rather than in $(r,r_\text{P})$.

To test whether our conclusions depend on the specific values of the constrained ratios we chose, we randomly perturb $s_\text{st}$, $K_\text{d}$, and $\mu$ and refit the free parameters $(r, r_\text{P})$ to obtain the best fit. Since $s_\text{st}$ and $K_\text{d}$ are ratios between rates and must remain positive, we multiply each by the exponential of a normal random variable with mean 0 and standard deviation 0.5. Since $\mu$ is the log ratio of rates, we instead add the normal random variable to $\mu$ directly. We further relax the assumptions that $\gamma_\text{ex}^\text{AB}=\gamma_\text{ex}^\text{CA}$ and that $\mu$ is the same across each pair of driven transitions. Instead, we perturb $s_\text{st}$ independently for $\gamma_\text{ex}^\text{AB}$ and $\gamma_\text{ex}^\text{CA}$, and perturb $\mu$ independently for each of the five transition pairs. Given the perturbed transition rates, we refit the parameters to minimize the KS statistic and extract the p-value. This process is repeated for 100 random trials. The p-value remains above 0.05 for 97 out of the 100 trials, indicating that the quality of the fit does not depend sensitively on the constrained parameters. Based on the same perturbation scheme, we further recompute the critical exponent $n^*$ in Sec. \ref{sec_analytical}, above which we obtain a peaked catastrophe length distribution (see Appendix \ref{appendix_peaked}). We repeat the calculation for 1000 trials. The exponent ranges from 0.09 to 0.4 and is concentrated within the range $[0.09,0.14]$ for 90\% of the trials.

\section{Condition for a peaked $P(x)$ distribution} \label{appendix_peaked}
In this section, we obtain an analytical condition for a peaked $P(x)$, which occurs when $P'(x^*)=0$ at $x^*>1$. We first derive the condition in terms of the bulk entry probability $p(x)$. Since $p(x)$ can be expressed in terms of transition rates, constraints on $p(x)$ further impose constraints on the GTP-tubulin dissociation rate.

First, we note that $p(x)$ must increase with $x$. To see this, we rewrite Equation (\ref{eqn_Px}) as the recurrence relation 
\begin{equation} \label{eqn_recur}
    P(x+1) = \frac{[1-p(x)]p(x+1)}{p(x)}P(x).
\end{equation}
When $p(x)$ decreases with $x$, i.e., $p(x+1)<p(x)$, the prefactor always satisfies $\frac{[1-p(x)]p(x+1)}{p(x)}<1$, which means that $P(x)$ decreases monotonically and there is no peak. Conversely, if $p(x)$ increases with $x$, the prefactor may exceed 1 and allows a peak to emerge. This is consistent with our model, where bulk entry is more likely for longer caps. While a peak can appear if $p(x)$ only increases within a limited range, we restrict the discussion to $p(x)$ that monotonically increases with $x$.

An increasing $p(x)$, however, is not sufficient for a peaked $P(x)$. As we can see from Equation (\ref{eqn_Px}), $p(x)$ must also increase sufficiently fast. To quantify this statement, we expand $P(x+1)$ to linear order
\begin{equation}
    P(x+1) \approx P(x) + P'(x).
\end{equation}
Here we assume that $P(x)$ varies slowly with $x$ and ignore higher order terms. Meanwhile, we obtain an equivalent expression by expanding $p(x+1)$ to linear order in Equation (\ref{eqn_recur}):
\begin{equation}
    P(x+1) \approx P(x)\frac{1-p(x)}{p(x)}\left[p(x)+p'(x)\right].
\end{equation}
Equating the two expressions and rearranging, we obtain
\begin{equation}
    \left[\log{P(x)}\right]'\approx \frac{1-p(x)}{p(x)} \left[p(x)+p'(x)\right] - 1.
\end{equation}
If $P(x)$ is peaked at some cap length $x^*$, we need $\left[\log{P(x^*)}\right]'=0$, i.e., 
\begin{equation}
    \frac{1-p(x^*)}{p(x^*)} \left[p(x^*)+p'(x^*)\right] - 1=0.
\end{equation}
This can be rewritten as
\begin{equation} \label{eqn_xmax}
    p'(x^*)=\frac{\left[p(x^*)\right]^2}{1-p(x^*)}.
\end{equation}
It follows that $P(x)$ is peaked if Equation (\ref{eqn_xmax}) has a solution at $x^*>1$.

To simplify Equation (\ref{eqn_xmax}), we further assume that $p(x)$ is a concave function of $x$. This assumption incurs minimal loss of generality, as an increasing probability function is typically concave down when it approaches its upper bound of 1. Indeed, the family of $p(x)$ functions we henceforth consider (general form given by Eq. (\ref{eqn_px})) are concave given our model parameters. Concavity means that $p'(x)$ on the left hand side of (\ref{eqn_xmax}) decreases with $x$. Meanwhile, the right hand side $\frac{\left[p(x)\right]^2}{1-p(x)}$ increases with $x$. Therefore, to obtain a solution satisfying $x^*>1$ for Equation (\ref{eqn_xmax}), we only require
\begin{equation} \label{eqn_cond}
    p'(1) > \frac{\left[p(1)\right]^2}{1-p(1)}.
\end{equation}
Using the approximation $p(2)=p(1)+p'(1)$, we can rewrite Equation (\ref{eqn_cond}) as
\begin{equation} \label{eqn_cond2}
    p(2) > \frac{p(1)}{1-p(1)}.
\end{equation}
This condition ensures a peaked $P(x)$ for an increasing concave function $p(x)$.

From Equation (\ref{eqn_cond2}), we obtain a lower bound on the GTP-tubulin dissociation rate $\gamma_\text{ex}^\text{CB}(l)$. Suppose that the dissociation rate grows with a power law: $\gamma_\text{ex}^\text{CB}(l)=\gamma_\text{ex}^\text{CB}(l=1) l^n$. Equation (\ref{eqn_cond2}) indicates that $p(x)$ and thus $\gamma_\text{ex}^\text{CB}(l)$ must increase sufficiently fast at $l=x=1$, captured by a lower bound on the exponent $n$. We solve for the lower bound numerically by substituting $p(x)$ (from Sec. \ref{sec_analytical}) and our model parameters (from Table \ref{table_parameters}) into Equation (\ref{eqn_cond2}), which yields $n\geq0.1$, consistent with our results in Fig. \ref{fig_analytical}(e). In other words, if the dissociation rate $\gamma_\text{ex}^\text{CB}\propto l^n$, it must grow faster than $l^{0.1}$ to produce a peaked $P(x)$ distribution.
\\
\section{Rescue events and critical concentration} \label{appendix_rescue}
In this section, we examine the average microtubule behavior in our model when rescue events are taken into account. Specifically, we show unbounded microtubule growth above some critical tubulin concentration, consistent with experimental observations \cite{walker1988dynamic,fygenson1994phase}.

Consider how the microtubule length changes in one growth-shortening cycle. The average length added in each growth episode equals the average catastrophe length, denoted by $L$. When rescue is possible, the average length removed in each shortening episode equals $v_\text{s} \bar{t}_{\text{re}}$, where $v_\text{s}$ is the shortening rate and $\bar{t}_{\text{re}}$ is the average time between catastrophe and regrowth. Regrowth can occur either through rescue or after complete microtubule depolymerization. When $v_\text{s} \bar{t}_{\text{re}}<L$, growth is unbounded because each shortening phase removes fewer tubulin subunits than each growth phase adds. At higher [Tub], $L$ increases in our model (Fig. \ref{fig_results}(c)), while $\bar{t}_{\text{res}}$ decreases and $v_\text{s}$ remains roughly constant based on experimental observations \cite{walker1988dynamic}. Because $v_\text{s} \bar{t}_{\text{res}}$ decreases while $L$ increases with increasing [Tub], there is a critical concentration over which $v_\text{s} \bar{t}_{\text{re}}<L$ and microtubules grow without bound.

\end{appendices}

\bibliographystyle{apsrev4-2}
\bibliography{references}% Produces the bibliography via BibTeX.

\end{document}